\documentclass[12pt,english]{article}
\usepackage[T1]{fontenc}
\usepackage[latin1]{inputenc}
\usepackage{a4wide}
\usepackage{babel}
\usepackage{graphics}
\usepackage{subfigure}

\makeatletter

\providecommand{\LyX}{L\kern-.1667em\lower.25em\hbox{Y}\kern-.125emX\@}
\newcommand{\noun}[1]{\textsc{#1}}

\usepackage{amssymb}

\makeatother

\begin{document}

\title{{\normalsize \begin{flushright}\normalsize{ITP--Budapest Report 583}\end{flushright}\vspace{1cm}}Spectrum
of boundary states in N=1 SUSY sine-Gordon theory}

\author{Z. Bajnok\thanks{
bajnok@afavant.elte.hu
}, L. Palla\thanks{
palla@ludens.elte.hu
}\ \ and G. Takács\thanks{
takacs@ludens.elte.hu
}\\
\\
{\normalsize Institute for Theoretical Physics}\\
{\normalsize Eötvös University}\\
{\normalsize H-1117 Budapest, Pázmány Péter sétány 1/A}\normalsize }

\date{25 June 2002}

\maketitle
\begin{abstract}
We consider N=1 supersymmetric sine-Gordon theory (\( SSG \)) with supersymmetric
integrable boundary conditions (boundary \( SSG \)=\( BSSG \)). We find two
possible ways to close the boundary bootstrap for this model, corresponding
to two different choices for the boundary supercharge.  
%conformal boundary conditions for the fermion field at
%the ultraviolet limit. 
We argue that these two bootstrap solutions should correspond
to the two integrable Lagrangian boundary theories considered recently by Nepomechie. 
\end{abstract}

\section{Introduction}

In this paper we consider the N=1 supersymmetric sine-Gordon theory with supersymmetric
integrable boundary conditions (\( BSSG \)). Our aim is to find a closure of
the boundary bootstrap for this model.

The N=1 supersymmetric sine-Gordon model is the natural supersymmetric extension
of the ordinary sine-Gordon model. It is an integrable field theory with infinitely
many conserved charges \cite{ferrara}. The \( S \) matrix of the theory was
obtained in \cite{ahn}, while the integrable and supersymmetric boundary conditions
were considered in \cite{inami}, but it took a while until the most general
integrable supersymmetric boundary interaction was found \cite{nepomechie}.

As a result of integrability, the boundary scattering factorizes, and the general
solution of the boundary Yang-Baxter equation was found in \cite{ahn_koo},
but the constraint of supersymmetry was not imposed. Nepomechie was the first
to consider supersymmetric boundary scattering \cite{nepomechie}, building
on previous results obtained in the case of supersymmetric sinh-Gordon theory
\cite{ahn_nepomechie}. However, no one has exposed the full structure of the
solitonic reflection amplitude, although the results obtained in the case of
the tricritical Ising model \cite{tricritical_boundary} are closely related
to this problem. Besides that, the closure of the bootstrap and the spectrum
of boundary states have not been even touched before. Therefore our aim is to
clear up the issue of supersymmetric boundary scattering in the \( BSSG \)
model and to find the complete spectrum of boundary states and their associated
reflection factors. The main idea -- motivated by the successful description
of the bulk scattering -- is to look for the reflection amplitudes in a form
where there is no mixing between the supersymmetric and other internal quantum
numbers. This means an Ansatz for the reflection amplitudes as a product of
two terms one of which is the ordinary (bosonic) sine-Gordon reflection amplitude,
while the other describes the scattering of the SUSY degrees of freedom.

Within the bootstrap procedure, we consider first the solitonic
reflection amplitudes on the ground state and on the first two excited
boundaries. The SUSY factors in these solutions have no poles in the
physical strip, thus the masses of boundary states emerging are the
same as in the bosonic theory, however, SUSY introduces a nontrivial
degeneracy. The spectrum of general higher excited boundaries is
easily extracted from these results. We determine also the various
reflection amplitudes on these excited boundaries.

There are two ways to close the bootstrap, starting from two different ground
state reflection amplitudes, corresponding to two possible choices of the boundary
supercharge. Both solutions lead to the same spectrum of boundary states, but
the reflection amplitudes are different. In one case the reflections
conserve fermionic parity, while in the other they do not.

The layout of the paper is as follows. Section 2 recalls briefly some important
facts about supersymmetric sine-Gordon theory. In Section 3 the reader is reminded
of the supersymmetric and integrable boundary interactions that can be added
to the theory, and we also discuss of the boundary supercharge and derive a
formula relating it to the boundary Hamiltonian. Section 4 gives a quick review
(containing only the most necessary facts) of the spectrum and reflection factors
of the ordinary (non supersymmetric) sine-Gordon model with integrable boundary
conditions. In Section 5 we present the main results of the paper, which is
a conjecture for the spectrum and the full set of reflection factors of the
\( BSSG \) model. We consider the breather reflection amplitudes in Section
6 and then give our conclusions in Section 7.

\section{Bulk SSG theory}

The bulk supersymmetric sine-Gordon theory is defined by the classical action\begin{eqnarray}
\mathcal{A}_{SSG} & = & \int dtdx\, \mathcal{L}_{SSG}(x,t)\nonumber \\
\mathcal{L}_{SSG} & = & \frac{1}{2}\partial _{\mu }\Phi \partial ^{\mu }\Phi +i\bar{\Psi }\gamma ^{\mu }\partial _{\mu }\Psi +m\bar{\Psi }\Psi \cos \frac{\beta }{2}\Phi +\frac{m^{2}}{\beta ^{2}}\cos \beta \Phi \label{bulk_action} 
\end{eqnarray}
where \( \Phi  \) is a real scalar, \( \Psi  \) is a Majorana fermion field,
\( m \) is a mass parameter and \( \beta  \) is the coupling constant. The
theory is invariant under an \( N=1 \) supersymmetry algebra and has infinitely
many commuting local conserved charges \cite{ferrara}. These charges survive
at the quantum level and render the theory integrable, which makes it possible
to describe the exact spectrum and the \( S \) matrix.

\subsection{Spectrum and bulk scattering amplitudes}

The spectrum consists of the soliton/antisoliton multiplet, realizing supersymmetry
in a nonlocal way, and breathers that are bound states of a soliton with an
antisoliton. 

The building blocks of supersymmetric factorized scattering theory were first
described in \cite{schoutens}, using an Ansatz in which the full scattering
amplitude is a direct product of a part carrying the SUSY structures and a part
describing all the rest of the dynamics. The full \( SSG \) \( S \) matrix
was constructed in \cite{ahn}.

The supersymmetric solitons are described by RSOS kinks \( K^{\epsilon }_{ab}\left( \theta \right)  \)
of mass \( M \) and rapidity \( \theta  \), where \( a,\: b \) take the values
\( 0,\frac{1}{2} \) and \( 1 \) with \( |a-b|=1/2 \), and describe the supersymmetric
structure, while \( \epsilon =\pm  \) corresponds to topological charge \( \pm 1 \)
(soliton/antisoliton). Multi-particle asymptotic states are built as follows\begin{equation}
\label{rsos_sequence}
\left| K^{\epsilon _{1}}_{a_{0}a_{1}}\left( \theta _{1}\right) K^{\epsilon _{2}}_{a_{1}a_{2}}\left( \theta _{2}\right) \dots K^{\epsilon _{N}}_{a_{N-1}a_{N}}\left( \theta _{N}\right) \right\rangle 
\end{equation}
where \( \theta _{1}>\theta _{2}>\dots >\theta _{N} \) for an \emph{in} state
and \( \theta _{1}<\theta _{2}<\dots <\theta _{N} \) \emph{}for an \emph{out}
state. The two-particle scattering process \[
K^{\epsilon _{1}}_{ab}\left( \theta _{1}\right) +K^{\epsilon _{2}}_{bc}\left( \theta _{2}\right) \, \rightarrow \, K^{\epsilon '_{2}}_{ad}\left( \theta _{2}\right) +K^{\epsilon '_{1}}_{dc}\left( \theta _{1}\right) \]
has an amplitude of the form\begin{equation}
\label{bulk_ampl}
S_{SUSY}\left( \left. \begin{array}{cc}
a & d\\
b & c
\end{array}\right| \theta _{1}-\theta _{2}\right) \times S_{SG}\left( \theta _{1}-\theta _{2}\right) ^{\epsilon '_{1}\epsilon '_{2}}_{\epsilon _{1}\epsilon _{2}}
\end{equation}
i.e. the tensor structure of the scattering amplitude factorizes into a part
describing the SUSY structure (which we call the SUSY factor) and another part
corresponding to the topological charge (the bosonic factor).

The bosonic factor coincides with the usual sine-Gordon \( S \) matrix\footnote{%
Note that the relation between the parameter \( \lambda  \) and the coupling
\( \beta  \) is different from the sine-Gordon case.
}\begin{eqnarray}
 & S_{SG}(u)_{++}^{++}=S_{SG}(u)^{--}_{--}= & -\prod ^{\infty }_{l=1}\left[ \frac{\Gamma (2(l-1)\lambda -\frac{\lambda u}{\pi })\Gamma (2l\lambda +1-\frac{\lambda u}{\pi })}{\Gamma ((2l-1)\lambda -\frac{\lambda u}{\pi })\Gamma ((2l-1)\lambda +1-\frac{\lambda u}{\pi })}/(u\rightarrow -u)\right] \nonumber \\
 & S_{SG}(u)^{+-}_{+-}=S_{SG}(u)^{-+}_{-+}= & \frac{\sin (\lambda u)}{\sin (\lambda (\pi -u))}S_{SG}(u)_{++}^{++}\quad \qquad ;\qquad \lambda =\frac{2\pi }{\beta ^{2}}-\frac{1}{2}\nonumber \\
 & S_{SG}(u)^{-+}_{+-}=S_{SG}(u)^{+-}_{-+}= & \frac{\sin (\lambda \pi )}{\sin (\lambda (\pi -u))}S_{SG}(u)_{++}^{++}\quad \qquad ;\qquad u=-i\theta \, \, \, ,\label{abc} 
\end{eqnarray}
while the SUSY factor is identical to the \( S \) matrix of the tricritical
Ising model perturbed by the primary field of dimension \( \frac{3}{5} \) \cite{tricritical}:\begin{eqnarray*}
S_{SUSY}\left( \left. \begin{array}{cc}
0 & \frac{1}{2}\\
\frac{1}{2} & 0
\end{array}\right| \theta \right) = & S_{SUSY}\left( \left. \begin{array}{cc}
1 & \frac{1}{2}\\
\frac{1}{2} & 1
\end{array}\right| \theta \right) = & 2^{(i\pi -\theta )/2\pi i}\cos \left( \frac{\theta }{4i}-\frac{\pi }{4}\right) K(\theta )\\
S_{SUSY}\left( \left. \begin{array}{cc}
\frac{1}{2} & 0\\
0 & \frac{1}{2}
\end{array}\right| \theta \right) = & S_{SUSY}\left( \left. \begin{array}{cc}
\frac{1}{2} & 1\\
1 & \frac{1}{2}
\end{array}\right| \theta \right) = & 2^{\theta /2\pi i}\cos \left( \frac{\theta }{4i}\right) K(\theta )\\
S_{SUSY}\left( \left. \begin{array}{cc}
0 & \frac{1}{2}\\
\frac{1}{2} & 1
\end{array}\right| \theta \right) = & S_{SUSY}\left( \left. \begin{array}{cc}
1 & \frac{1}{2}\\
\frac{1}{2} & 0
\end{array}\right| \theta \right) = & 2^{(i\pi -\theta )/2\pi i}\cos \left( \frac{\theta }{4i}+\frac{\pi }{4}\right) K(\theta )\\
S_{SUSY}\left( \left. \begin{array}{cc}
\frac{1}{2} & 1\\
0 & \frac{1}{2}
\end{array}\right| \theta \right) = & S_{SUSY}\left( \left. \begin{array}{cc}
\frac{1}{2} & 0\\
1 & \frac{1}{2}
\end{array}\right| \theta \right) = & 2^{\theta /2\pi i}\cos \left( \frac{\theta }{4i}-\frac{\pi }{2}\right) K(\theta )
\end{eqnarray*}
\[
K(\theta )=\frac{1}{\sqrt{\pi }}\prod ^{\infty }_{k=1}\frac{\Gamma \left( k-1/2+\theta /2\pi i\right) \Gamma \left( k-\theta /2\pi i\right) }{\Gamma \left( k+1/2-\theta /2\pi i\right) \Gamma \left( k+\theta /2\pi i\right) }\]
As the SUSY factor has no poles in the physical strip, the solitonic amplitudes
(\ref{bulk_ampl}) have poles at exactly the same locations as the sine-Gordon
soliton \( S \) matrix. These correspond to bound states (breathers) \( \mathbf{B}_{n} \)
of mass\[
m_{n}=2M\sin \frac{\pi n}{2\lambda }\: ,\: n=1,\dots \, [\lambda ]\,
.\]
The $S$ matrix of the breathers was first found in \cite{shankar}.
The breathers form a particle multiplet composed of a boson and a fermion, on
which supersymmetry is represented in a standard way \cite{schoutens,hollowood}.

For the ordinary sine-Gordon theory, the correspondence between the Lagrangian
theory and the bootstrap \( S \) matrix (\ref{abc}) is very well established.
There is much less evidence for the correctness of the \( S \) matrix (\ref{bulk_ampl})
as the scattering amplitude of SUSY sine-Gordon theory. Besides the original
construction \cite{ahn} (based on arguments related to \( N=1 \) supersymmetric
minimal models), another indication is that at a particular value of the coupling
\( \beta  \) where it is expected to have a restriction to the SUSY version
of Lee-Yang theory (superconformal minimal model \( \mathcal{SM}\left( 2/8\right)  \)
perturbed by the relevant superconformal primary field \( \Phi \left( 1,3\right)  \),
which is equivalent to Virasoro minimal model \( \mathcal{M}\left( 3/8\right)  \)
perturbed by the primary field \( \Phi \left( 1,5\right)  \)),
the first breather supermultiplet has the same scattering amplitude as predicted
from RSOS restriction of imaginary coupled \( a^{(2)}_{2} \) Toda
theory in \cite{phi15} (see also \cite{sleeyang}).

\subsection{Bulk SUSY charges}

The bulk theory has two supersymmetry charges of opposite chirality \( Q \)
and \( \bar{Q} \), which together form a Majorana spinor. They act on one-particle
states \( \left| A_{i}(\theta )\right\rangle  \) in the following way \cite{tricritical,schoutens,hollowood}:\[
Q\left| A_{i}(\theta )\right\rangle =\sqrt{m_{i}}\mathrm{e}^{\theta /2}\mathcal{Q}\left| A_{i}(\theta )\right\rangle \quad ,\quad \bar{Q}\left| A_{i}(\theta )\right\rangle =\sqrt{m_{i}}\mathrm{e}^{-\theta /2}\bar{\mathcal{Q}}\left| A_{i}(\theta )\right\rangle \]
where \( m_{i} \) are the particle masses and \( \mathcal{Q},\: \bar{\mathcal{Q}} \)
are matrices satisfying \[
\mathcal{Q}^{2}=1\quad ,\quad \bar{\mathcal{Q}}^{2}=1\; .\]
In the one-particle basis \( \left\{ K_{0\frac{1}{2}}\, ,K_{1\frac{1}{2}}\, ,K_{\frac{1}{2}0}\, ,K_{\frac{1}{2}1}\right\}  \)
(we omit the upper index \( \epsilon  \), as the SUSY action does not depend
on the topological charge), the supersymmetry algebra is represented by the
matrices\[
\mathcal{Q}=\left( \begin{array}{cccc}
0 & i & 0 & 0\\
-i & 0 & 0 & 0\\
0 & 0 & 1 & 0\\
0 & 0 & 0 & -1
\end{array}\right) \qquad ,\qquad \bar{\mathcal{Q}}=\left( \begin{array}{cccc}
0 & i & 0 & 0\\
-i & 0 & 0 & 0\\
0 & 0 & -1 & 0\\
0 & 0 & 0 & 1
\end{array}\right) \]
The SUSY algebra of the sine-Gordon theory has a central charge and the matrix\[
Z=\frac{1}{2}\left\{ \mathcal{Q},\bar{\mathcal{Q}}\right\} =\left( \begin{array}{cccc}
1 & 0 & 0 & 0\\
0 & 1 & 0 & 0\\
0 & 0 & -1 & 0\\
0 & 0 & 0 & -1
\end{array}\right) \]
describes the SUSY central charge in the above basis. This is not to be confused
with the topological charge \( T \) of the sine-Gordon solitons, which is represented
by the upper indices \( \epsilon  \). \( Z \) can take the values \( 0 \)
or \( \pm 1 \), and it distinguishes between solitons/antisolitons mediating
from odd to even and from even to odd vacua of the bosonic potential \cite{witten_olive}
(this means that using the terminology of \cite{kfolded} the theory is 2-folded). 

The above representation of SUSY describes BPS saturated objects. There was
a controversy in the literature whether the solitons in \( SSG \) are BPS saturated,
since \( N=1 \) SUSY does not protect their mass from acquiring radiative corrections.
However, it was shown in \cite{bps} that they remain BPS saturated at one-loop
and probably to all orders, due to anomalous quantum corrections to the classical
formula for the central charge \( Z \).

The fermionic parity operator \( \Gamma =(-1)^{F} \) is given by the matrix\[
\Gamma =\left( \begin{array}{cccc}
0 & 1 & 0 & 0\\
1 & 0 & 0 & 0\\
0 & 0 & 0 & 1\\
0 & 0 & 1 & 0
\end{array}\right) \]
Using the definition of \( \Gamma  \) it is possible to specify a basis of
pure bosonic and pure fermionic states for any given (fixed) number of particles.
However, the composition (coproduct) rules of the kink states as given in (\ref{rsos_sequence})
are not free and therefore in the boson-fermion internal space supersymmetry
acts nonlocally. The action of supersymmetry on multi-particle states involves
braiding factors depending on \( \Gamma  \) that are defined by the coproduct
\( \Delta : \)\begin{eqnarray*}
\Delta \left( Q\right)  & = & Q\otimes \mathbb {I}+\Gamma \otimes Q\\
\Delta \left( \bar{Q}\right)  & = & \bar{Q}\otimes \mathbb {I}+\Gamma \otimes \bar{Q}\\
\Delta \left( \Gamma \right)  & = & \Gamma \otimes \Gamma 
\end{eqnarray*}
The action on breather states can be derived using the bootstrap, but can also
be obtained from the representation theory of the SUSY algebra. It turns out
that the central charge \( Z \) (as well as the topological charge \( T \))
vanishes identically for the breathers. For further details we refer to \cite{schoutens,hollowood}.

\section{Boundary SSG theory}

\subsection{Integrable and supersymmetric boundary interactions}

Boundary \( SSG \) (\( BSSG \)) theory can be described by an action of the
form\[
\mathcal{A}_{BSSG}=\int ^{\infty }_{-\infty }dt\int ^{0}_{-\infty }dx\mathcal{L}_{SSG}(x,t)+\int ^{\infty }_{-\infty }L_{B}(t)dt\]
where \( L_{B}(t) \) is a term local at \( x=0 \). We consider boundary interactions
which are both supersymmetric and integrable. The case when \( L_{B}(t) \)
depends only on the value of \( \Phi  \) and \( \Psi  \) at \( x=0 \) was
considered in \cite{inami}, where it was found that supersymmetry and integrability
restricts the boundary interaction to the form\begin{equation}
\label{boundary_interaction_inami}
L_{B}=\left. \pm \frac{4m}{\beta ^{2}}\cos \frac{\beta }{2}\Phi \pm \bar{\psi }\psi \right| _{x=0}
\end{equation}
which gives four discrete choices. Here we write the Majorana spinor in component
form\[
\Psi =\left( \begin{array}{c}
\bar{\psi }\\
\psi 
\end{array}\right) \; .\]

However, this is not the whole story. The Majorana fermion in the ultraviolet
limit is described by the \( c=\frac{1}{2} \) Ising conformal field theory.
It was noted in \cite{chatterjee} that in order for the Majorana fermion to
describe correctly the boundary states of the conformal field theory, one has
to include a fermionic boundary degree of freedom \( a(t) \) that is related
to the boundary value of the Ising spin. Using this fact, a two-parameter set
of integrable supersymmetric boundary conditions was derived in \cite{nepomechie}.
The boundary interaction term is of the form\begin{equation}
\label{boundary_interaction}
L^{\pm }_{B}=\left. \left( \pm \bar{\psi }\psi +ia\partial _{t}a-2f\left( \Phi \right) a(\psi \mp \bar{\psi })+\mathcal{B}\left( \Phi \right) \right) \right| _{x=0}
\end{equation}
The functions \( f \) and \( \mathcal{B} \) are fixed by the requirement of
boundary integrability and supersymmetry (cf. \cite{nepomechie}). We denote
the theories obtained by adding \( L^{\pm }_{B} \) to the bulk action as \( BSSG^{\pm } \).
The most important property of this action is that it depends on two continuously
varying boundary parameters, exactly as in the case of the non-supersymmetric
boundary sine-Gordon theory \cite{GZ}. This is important for consistency with
the bootstrap since the reflection factors we find depend on two parameters
as well. The boundary Lagrangian (\ref{boundary_interaction_inami}) can be
obtained as a special case of (\ref{boundary_interaction}) when the parameters
are tuned so that \( f=0 \), and therefore the boundary fermion \( a \) decouples.

\subsection{The boundary SUSY charge and the Hamiltonian}

The bulk SUSY charges \( Q \) and \( \bar{Q} \) can be written as integrals
of local (fermionic) densities \( q \) and \( \bar{q} \):\[
Q=\int ^{\infty }_{-\infty }q(x,t)dx\quad ,\quad \bar{Q}=\int ^{\infty }_{-\infty }\bar{q}(x,t)dx\]
They have the anticommutation relation
\[
\{Q,\bar{Q}\}=2M\, Z\]
and satisfy the following relation (among others)\begin{equation}
\label{q2bulk}
QQ+\bar{Q}\bar{Q}=2H
\end{equation}
 where \( H=\int h(x,t)dx \) is the Hamiltonian, \( Z \) is the SUSY central
charge and \( M \) is the soliton mass. In a boundary theory with supersymmetric
integrable boundary condition, the conserved supercharge can be written as follows:\[
\tilde{Q}_{\pm }=\int ^{0}_{-\infty }\left( q(x,t)\pm \bar{q}(x,t)\right) dx+Q_{B}(x=0,t)\]
where \( Q_{B} \) is a boundary contribution, localized at \( x=0 \). There
are two possible choices (\( \pm  \)) corresponding to the two possible conformal
fermionic boundary conditions in the ultraviolet limit. As shown in \cite{nepomechie},
these correspond to the two choices of sign in (\ref{boundary_interaction})
and therefore to \( BSSG^{\pm } \). 

Similarly, the Hamiltonian takes the form\[
\tilde{H}=\int ^{0}_{-\infty }h(x,t)dx+H_{B}(x=0,t)\]
where \( H_{B} \) is the boundary interaction. Let us for the moment neglect
the contribution from the central charge \( Z \) (this is possible e.g. in
a sector containing only breathers). Then, using eqn. (\ref{q2bulk}) it is
easy to see that\[
\tilde{Q}^{2}_{\pm }=2\int ^{0}_{-\infty }h(x,t)dx+2H_{B}^{\prime }(x=0,t)\]
where \( H_{B}^{\prime } \) is again some term localized at \( x=0 \). However,
\( \tilde{Q}_{\pm } \) is conserved in time, and therefore\[
\frac{d}{dt}\left( 2\int ^{0}_{-\infty }h(x,t)dx+2H_{B}^{\prime }(x=0,t)\right) =0.\]
In the supersymmetric boundary sine-Gordon theory, this property uniquely determines
\( H_{B}^{\prime } \) as the boundary interaction \( H_{B} \) and therefore\[
\tilde{Q}^{2}_{\pm }=2\tilde{H}\]
Including \( Z \), it is natural to expect that this relation extends as follows:\begin{equation}
\label{q2relation}
\tilde{Q}^{2}_{\pm }=2\left( \tilde{H}\pm M\, \tilde{Z}\right) 
\end{equation}
where \( \tilde{Z} \) is an appropriate extension of \( Z \) to the boundary
situation. We shall see that the two bootstrap solutions we propose correctly
reproduce this formula.

\section{Boundary sine-Gordon model}

\subsection{Ground state reflection factors}

The most general reflection factor -- modulo CDD-type factors -- of the soliton
antisoliton multiplet \( |s,\bar{s}\rangle  \) on the ground state boundary,
denoted by \( |\, \rangle  \), satisfying the boundary versions of the Yang-Baxter,
unitarity and crossing equations was found by Ghoshal and Zamolodchikov \cite{GZ}:\begin{eqnarray}
R_{SG}(\eta ,\vartheta ,u) & = & \left( \begin{array}{cc}
P^{+}(\eta ,\vartheta ,u) & Q(\eta ,\vartheta ,u)\\
Q(\eta ,\vartheta ,u) & P^{-}(\eta ,\vartheta ,u)
\end{array}\right) \nonumber \\
 & = & \left( \begin{array}{cc}
P_{0}^{+}(\eta ,\vartheta ,u) & Q_{0}(u)\\
Q_{0}(u) & P_{0}^{-}(\eta ,\vartheta ,u)
\end{array}\right) R_{0}(u)\frac{\sigma (\eta ,u)}{\cos (\eta )}\frac{\sigma (i\vartheta ,u)}{\cosh (\vartheta )}\quad ,\nonumber \\
P_{0}^{\pm }(\eta ,\vartheta ,u) & = & \cos (\lambda u)\cos (\eta )\cosh (\vartheta )\mp \sin (\lambda u)\sin (\eta )\sinh (\vartheta )\quad ,\nonumber \\
Q_{0}(u) & = & -\sin (\lambda u)\cos (\lambda u)\quad ,\label{Rsas} 
\end{eqnarray}
 where \( \eta  \) and \( \vartheta  \) are the two real parameters characterizing
the solution, \[
R_{0}(u)=\prod ^{\infty }_{l=1}\left[ \frac{\Gamma \left( 4l\lambda -\frac{2\lambda u}{\pi }\right) \Gamma \left( 4\lambda (l-1)+1-\frac{2\lambda u}{\pi }\right) }{\Gamma \left( (4l-3)\lambda -\frac{2\lambda u}{\pi }\right) \Gamma \left( (4l-1)\lambda +1-\frac{2\lambda u}{\pi }\right) }/(u\to -u)\right] \]
 is the boundary condition independent part and \[
\sigma (x,u)=\frac{\cos x}{\cos (x+\lambda u)}\prod ^{\infty }_{l=1}\left[ \frac{\Gamma \left( \frac{1}{2}+\frac{x}{\pi }+(2l-1)\lambda -\frac{\lambda u}{\pi }\right) \Gamma \left( \frac{1}{2}-\frac{x}{\pi }+(2l-1)\lambda -\frac{\lambda u}{\pi }\right) }{\Gamma \left( \frac{1}{2}-\frac{x}{\pi }+(2l-2)\lambda -\frac{\lambda u}{\pi }\right) \Gamma \left( \frac{1}{2}+\frac{x}{\pi }+2l\lambda -\frac{\lambda u}{\pi }\right) }/(u\to -u)\right] \]
describes the boundary condition dependence. The reflection factors of the breathers
can be obtained by the bulk bootstrap procedure \cite{bulk_bootstrap}.

\subsection{The general spectrum and the associated reflection factors}

The spectrum of boundary excited states was determined in \cite{dirichlet,BSG}.
It can be parametrized by a sequence of integers \( |n_{1},n_{2},\dots ,n_{k}\rangle  \),
whenever the \( \frac{\pi }{2}\geq \nu _{n_{1}}>w_{n_{2}}>\dots \geq 0 \) condition
holds, where \[
\nu _{n}=\frac{\eta }{\lambda }-\frac{\pi (2n+1)}{2\lambda }\quad \textrm{and}\qquad w_{k}=\pi -\frac{\eta }{\lambda }-\frac{\pi (2k-1)}{2\lambda }\; .\]
 The mass of such a state is 

\begin{equation}
\label{exc_energy}
m_{|n_{1},n_{2},\dots ,n_{k}\rangle }=M\sum _{i\textrm{ odd}}\cos (\nu _{n_{i}})+M\sum _{i\textrm{ even}}\cos (w_{n_{i}})\, \, \, .
\end{equation}
The reflection factors depend on which sectors we are considering. In the even
sector, i.e. when \( k \) is even, we have\[
Q_{|n_{1},n_{2},\dots ,n_{k}\rangle }(\eta ,\vartheta ,u)=Q(\eta ,\vartheta ,u)\prod _{i\textrm{ odd}}a_{n_{i}}(\eta ,u)\prod _{i\textrm{ even}}a_{n_{i}}(\bar{\eta },u)\, \, \, ,\]
and \[
P^{\pm }_{|n_{1},n_{2},\dots ,n_{k}\rangle }(\eta ,\vartheta ,u)=P^{\pm }(\eta ,\vartheta ,u)\prod _{i\textrm{ odd}}a_{n_{i}}(\eta ,u)\prod _{i\textrm{ even}}a_{n_{i}}(\bar{\eta },u)\, \, \, ,\]
for the solitonic processes, where \[
a_{n}(\eta ,u)=\prod _{l=1}^{n}\left\{ 2\left( \frac{\eta }{\pi }-l\right) \right\} \quad ;\qquad \bar{\eta }=\pi (\lambda +1)-\eta \]
and\[
\{y\}=\frac{\left( \frac{y+1}{2\lambda }\right) \left( \frac{y-1}{2\lambda }\right) }{\left( \frac{y+1}{2\lambda }-1\right) \left( \frac{y-1}{2\lambda }+1\right) }\quad ,\quad (x)=\frac{\sin \left( \frac{u}{2}+\frac{x\pi }{2}\right) }{\sin \left( \frac{u}{2}-\frac{x\pi }{2}\right) }\]
 In the odd sector, i.e. when \( k \) is odd, the same formulae apply if in
the ground state reflection factors the \( \eta \leftrightarrow \bar{\eta } \)
and \( s\leftrightarrow \bar{s} \) changes are made. The breather sector can
be obtained again by bulk fusion.

\section{Supersymmetric boundary sine-Gordon model}

\subsection{Ground state reflection factors}

\subsubsection{The general solution for the reflection factor}

Following the bulk case, we suppose that the reflection matrix factorizes as
\[
R_{SUSY}(\theta )\times R_{SG}(\theta )\; .\]
 In this special form the constraints as unitarity, boundary Yang-Baxter equation
and crossing-unitarity relation \cite{GZ} can be satisfied separately for the
two factors. Since the sine-Gordon part already fulfills these requirements,
we concentrate on the supersymmetric part. 

From the RSOS nature of the bulk \( S \)-matrix (\ref{bulk_ampl}) it is clear
that the boundary must also have RSOS labels and the adjacency conditions between
the nearest kink and the boundary must also hold. Thus the following reflections
are possible:\[
K_{ba}(\theta )|B_{a}\rangle =\sum _{c}R^{b}_{ac}(\theta )K_{bc}(-\theta )|B_{c}\rangle \]
or in detail\[
K_{0\frac{1}{2}}(\theta )|B_{\frac{1}{2}}\rangle =R^{0}_{\frac{1}{2}\frac{1}{2}}(\theta )K_{0\frac{1}{2}}(-\theta )|B_{\frac{1}{2}}\rangle \quad ;\qquad K_{1\frac{1}{2}}(\theta )|B_{\frac{1}{2}}\rangle =R^{1}_{\frac{1}{2}\frac{1}{2}}(\theta )K_{1\frac{1}{2}}(-\theta )|B_{\frac{1}{2}}\rangle \; .\]
 and \begin{equation}
\label{dubl}
K_{\frac{1}{2}a}(\theta )|B_{a}\rangle =R^{\frac{1}{2}}_{aa}(\theta )K_{\frac{1}{2}a}(-\theta )|B_{a}\rangle +R^{\frac{1}{2}}_{ab}(\theta )K_{\frac{1}{2}b}(-\theta )|B_{b}\rangle \quad ,\quad b\neq a\quad ,\quad a,b=0,1
\end{equation}
In the second process the label of the boundary state has changed, which shows
that \( |B_{0}\rangle  \) and \( |B_{1}\rangle  \) form a doublet. All of
the constraints mentioned above factorize in the sense that they give independent
equations for the reflections on the boundary \( |B_{1/2}\rangle  \) and on
the doublet \( |B_{0,1}\rangle  \). Since the ground state boundary is expected
to be nondegenerate we first concentrate on reflection factors off the singlet
boundary \( |B_{1/2}\rangle  \). The most general solution of the boundary
Yang-Baxter equation is of the form \cite{ahn_koo}\[
R^{0}_{\frac{1}{2}\frac{1}{2}}(\theta )=(1+A\sinh (\theta /2))M(\theta )\quad ;\qquad R^{1}_{\frac{1}{2}\frac{1}{2}}(\theta )=(1-A\sinh (\theta /2))M(\theta )\]
while unitarity and crossing symmetry give the following restrictions\begin{eqnarray*}
 &  & M(\theta )M(-\theta )(1-A^{2}\sinh ^{2}(\theta /2))=1\; ,\\
 &  & M\left( \frac{i\pi }{2}-\theta \right) =\cosh \left( \frac{\theta }{2}-\frac{i\pi }{4}\right) K(2\theta )2^{(i\pi +2\theta )/2\pi i}M\left( \frac{i\pi }{2}+\theta \right) \; .
\end{eqnarray*}
We suppose that the boundary states \( \left| B_{a}\right\rangle \: ,\: a=0,1 \)
can be obtained by boundary bootstrap from the ground state \( \left| B_{1/2}\right\rangle  \)
\cite{chim,tricritical_boundary}. Therefore we do not consider the Yang-Baxter
equation and the other constraints for the amplitudes \( R^{1/2}_{ab}\left( \theta \right)  \)
as these will be guaranteed to be fulfilled by the bootstrap. We shall see later
that in general the boundary states \( \left| B_{a}\right\rangle ,\: a=0,1 \)
come in multiple copies, each of which forms a doublet of states with the same
energy.

\subsubsection{Action of the boundary supersymmetry charges \protect\( \tilde{Q}_{\pm }\protect \)}

We need to construct the action of the boundary supercharge on the asymptotic
states. We expect that the action is given by \[
\tilde{Q}_{\pm }=Q\pm \bar{Q}+Q_{B}\]
where \( Q \), \( \bar{Q} \) act on the particles as in the bulk theory (Section
2.2)\footnote{%
Note that our charge \( \bar{Q} \) differs by a sign from that used in \cite{tricritical_boundary},
while agrees with the convention in \cite{hollowood}. This is important when
comparing our results with those in \cite{tricritical_boundary}.
}. The reason for this is that they are given by integrals of local (fermionic)
densities, and asymptotic particles are localized far away from the wall, so
the action of these charges is not affected by the presence of the boundary.
\( Q_{B} \) is the action of the boundary contribution, which we take to be
\begin{equation}
\label{QB}
Q_{B}=\gamma \Gamma \: ,
\end{equation}
where \( \gamma  \) is some unknown parameter (related to the energy of the
boundary ground state -- see later). The reason for this choice is that we expect
the boundary supercharge to commute with the bulk \( S \)-matrix, which is
symmetric under the action of \( Q,\bar{Q} \) and \( \Gamma  \) by construction\cite{schoutens,hollowood}.
(\ref{QB}) is also supported by classical considerations in
\cite{nepomechie}, showing also that the classical version of
\(\gamma\) is a function of the parameters in the boundary Lagrangian 
(\ref{boundary_interaction}).

Next we need to give the action of \( Q \), \( \bar{Q} \) and \( \Gamma  \)
on the boundary ground state \( \left| B_{1/2}\right\rangle  \). Following
\cite{tricritical_boundary}, we choose\begin{equation}
\label{gr_action}
Q|B_{\frac{1}{2}}\rangle =0\quad ,\quad \bar{Q}|B_{\frac{1}{2}}\rangle =0\quad ,\quad \Gamma |B_{\frac{1}{2}}\rangle =|B_{\frac{1}{2}}\rangle \; .
\end{equation}
 The first two relations express that the boundary ground state is supersymmetric,
while the last one shows that it is an eigenvector of \( \Gamma  \). We expect
that because the ground state is nondegenerate. The choice of the eigenvalue
(\( \pm 1 \)) is not important, as it could be compensated by a redefinition
of \( \gamma  \). It is a consequence of (\ref{q2relation}) and (\ref{gr_action})
that the ground state energy is \( \gamma ^{2}/2 \), which will be shown later
to be consistent with the action of \( \tilde{Q}_{\pm } \) on the excited boundary
states.

\subsubsection{Supersymmetric reflection amplitudes}

Now we would like to impose the supersymmetry constraints on the ground state
reflection amplitudes. 

The two choices \( \tilde{Q}_{\pm } \) will give different solutions \cite{tricritical_boundary,chim}.
If the boundary supercharge \( \tilde{Q}_{+} \) commutes with the reflections
(theory \( BSSG^{+} \)) then we obtain\begin{equation}
\label{Rplus}
R^{0}_{\frac{1}{2}\frac{1}{2}}(\theta )=R^{1}_{\frac{1}{2}\frac{1}{2}}(\theta )=2^{-\theta /\pi i}\prod _{k=1}^{\infty }\left[ \frac{\Gamma (k-\frac{\theta }{2\pi i})\Gamma (k-\frac{\theta }{2\pi i})}{\Gamma (k-\frac{1}{4}-\frac{\theta }{2\pi i})\Gamma (k+\frac{1}{4}-\frac{\theta }{2\pi i})}/\left\{ \theta \, \leftrightarrow -\theta \right\} \right] =2^{-\theta /\pi i}P(\theta )
\end{equation}
If, however, it is \( \tilde{Q}_{-} \) that commutes with the reflections (\( BSSG^{-} \))
then the result is\begin{eqnarray}
R^{0}_{\frac{1}{2}\frac{1}{2}}(\theta ) & = & \left( \cos \frac{\xi }{2}+i\sinh \frac{\theta }{2}\right) K(\theta -i\xi )K(i\pi -\theta -i\xi )2^{-\theta /\pi i}P(\theta )\nonumber \\
R^{1}_{\frac{1}{2}\frac{1}{2}}(\theta ) & = & \left( \cos \frac{\xi }{2}-i\sinh \frac{\theta }{2}\right) K(\theta -i\xi )K(i\pi -\theta -i\xi )2^{-\theta /\pi i}P(\theta )\label{Rminus} 
\end{eqnarray}
where \( \xi  \) is related to \( \gamma  \) as \begin{equation}
\label{gamma_xi}
\gamma =-2\sqrt{M}\cos \frac{\xi }{2}\: .
\end{equation}
Note that symmetry of the reflection under \( \Gamma  \) requires

\[
R_{\frac{1}{2}\frac{1}{2}}^{0}\left( \theta \right) =R_{\frac{1}{2}\frac{1}{2}}^{1}\left( \theta \right) \; ,\]
thus in the first case (\( BSSG^{+} \)) the reflections also commute with the
operator \( \Gamma  \), while in the other case (\( BSSG^{-} \)) they do not.
We remark that there are no poles in the physical strip in any of the reflection
factors above.

In the case \( BSSG^{+} \), the supersymmetry constraints do not determine
the value of \( \gamma  \) in contrast to the results of \cite{tricritical_boundary,chim}.
The reason is that it is the supersymmetry of the reflections on \( \left| B_{a}\right\rangle ,\: a=0,1 \)
which connects \( \gamma  \) with a parameter in the reflection matrix itself.
However, we construct these reflections by the bootstrap, which determines them
completely, and \( \gamma  \) is left as a free parameter. As it was
argued above and as will also be seen later 
\( \gamma  \) is connected to the vacuum energy, so it is not a new parameter
of the theory (in principle it is expressible in terms of the Lagrangian parameters).
The only independent parameters introduced by the boundary are \( \eta  \)
and \( \vartheta  \) which are present in the bosonic sine-Gordon reflection
factors \( R_{SG} \).

\subsection{The general spectrum and the associated reflection factors}

\subsubsection{The \protect\( \Gamma \protect \) symmetric case (\protect\( BSSG^{+}\protect \))}

We start with the analysis of the ground state reflection factors\[
R_{\frac{1}{2}\frac{1}{2}}^{a}\left( \theta \right) \times R_{SG}\left( \theta \right) \]
where the SUSY component has the form (\ref{Rplus}). Since the only poles of
these reflection factors are due to the sine-Gordon part their explanation has
to be similar to that in the bosonic theory. However, we have to supplement
the formulae for the bosonic theory with RSOS indices in a consistent way. The
sine-Gordon reflection factor has boundary independent poles at \( i\frac{n\pi }{2\lambda } \)
for \( n=1,2,\dots  \), which can be described by diagram (a). This is identical
to the non supersymmetric diagram except that it is decorated with RSOS indices,
which are displayed inside circles. Clearly the dashed line denotes the full breather
supermultiplet (now consisting of a boson and a fermion).

\begin{figure}\subfigure[Soliton bulk pole]{\resizebox*{!}{6cm}{\includegraphics{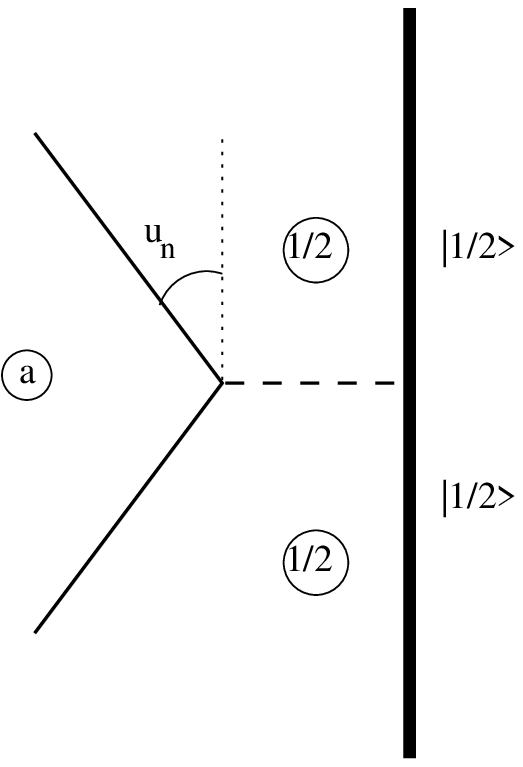}}} ~~~~~~~~~~~~~~~\subfigure[bootstrap I. ]{\resizebox*{!}{6cm}{\includegraphics{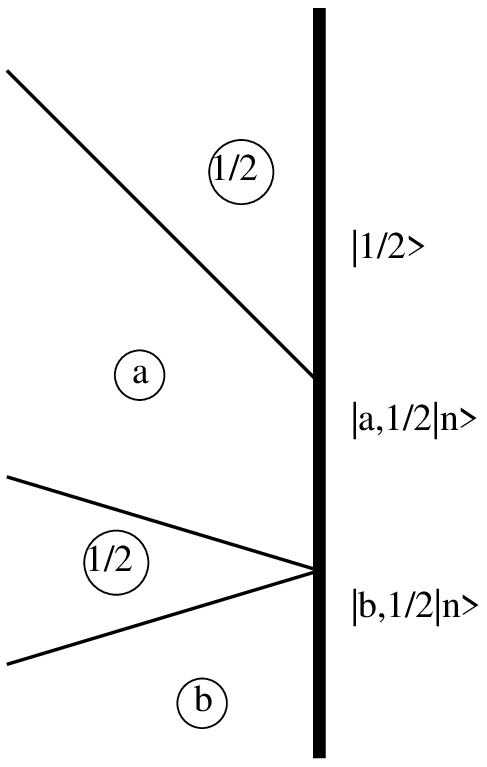}}} ~~~~~~~~~~~~~~~~\subfigure[bootstrap II.]{\resizebox*{!}{6cm}{\includegraphics{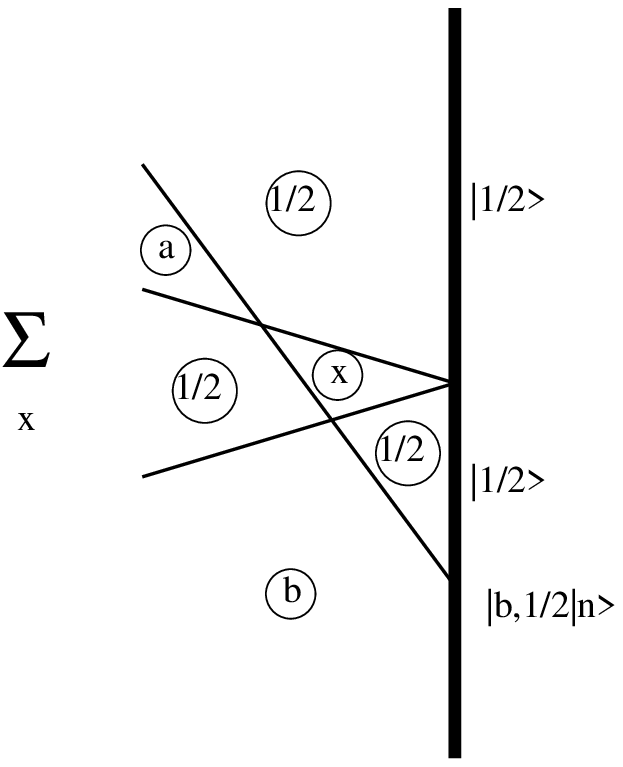}}} ~~~~~\end{figure}

\noindent The boundary dependent poles of \( R_{SG} \) are located at
\[ -i\theta =
\nu _{n}=\frac{\eta }{\lambda }-\frac{(2n+1)\pi }{2\lambda }\quad ;\qquad n=0,1,\dots \, \, \, .\]
 At the position of these poles we associate boundary bound states to the reflection
amplitudes \( R^{a}_{\frac{1}{2}\frac{1}{2}}\, ,\, a=0,1 \) \begin{equation}
\label{first_fusion}
\left| \left. a,1/2\right| n\right\rangle =\frac{1}{g^{\left| 1/2\right\rangle }_{\left| \left. a,1/2\right| n\right\rangle }}K_{a\frac{1}{2}}\left( i\, \nu _{n}\right) \left| \frac{1}{2}\right\rangle \; ,\quad \mathrm{where}\; \left| \frac{1}{2}\right\rangle \equiv \left| B_{\frac{1}{2}}\right\rangle \; ,
\end{equation}
where the \( g \)-factor is the SUSY part of the boundary coupling, coming
from the SUSY component of the reflection factor (for definitions of boundary
couplings, see \cite{GZ}). The two states (\( a=0,\, 1 \)) for a given \( n \)
form a doublet which realizes the structure (\ref{dubl}), that is the \( K_{\frac{1}{2}a} \)
kinks can scatter on it. The action of the boundary supercharge on these states
can be calculated using the coproduct rules in \cite{schoutens,hollowood},
taking into account the action of the charges on the boundary ground state:
\begin{eqnarray}
\tilde{Q}_{+}\left| \left. 0,1/2\right| n\right\rangle  & = & r\left( \gamma -2i\sqrt{M}\cos \frac{\nu _{n}}{2}\right) \left| \left. 1,1/2\right| n\right\rangle \nonumber \\
\tilde{Q}_{+}\left| \left. 1,1/2\right| n\right\rangle  & = & r^{-1}\left( \gamma +2i\sqrt{M}\cos \frac{\nu _{n}}{2}\right) \left| \left. 0,1/2\right| n\right\rangle \quad ,\quad r=\frac{g^{\left| 1/2\right\rangle }_{\left| \left. 1,1/2\right| n\right\rangle }}{g^{\left| 1/2\right\rangle }_{\left| \left. 0,1/2\right| n\right\rangle }}\label{excited_Q} 
\end{eqnarray}
The boundary supercharge satisfies\[
\tilde{Q}^{2}_{+}\left| \left. a,1/2\right| n\right\rangle =2\left( \frac{\gamma ^{2}}{2}+M\cos \nu _{n}+M\right) \left| \left. a,1/2\right| n\right\rangle \]
which is exactly the relation \( \tilde{Q}^{2}_{+}=2\left( \tilde{H}+M\, \tilde{Z}\right)  \),
since the central charge of this state is \( \tilde{Z}=1 \) (we take the ground
state \( \left| 1/2\right\rangle  \) to have \( \tilde{Z}=0 \), while the
bulk soliton \( K_{a1/2} \) has \( Z=1 \)), the ground state has energy \( \gamma ^{2}/2 \)
by virtue of the relations (\ref{QB},\ref{gr_action}) and \( M\cos \nu _{n} \)
is the energy that the excited state has relative to the ground state (\ref{exc_energy}).

The SUSY reflection factors of \( K_{\frac{1}{2}a} \) off \( \left| \left. a,1/2\right| n\right\rangle  \)
can be computed from the bootstrap principle (diagrams (b) and (c)):
\begin{eqnarray}
g^{\frac{1}{2}}_{a}R_{ab}^{\frac{1}{2}}(\theta ) & = & g_{b}^{\frac{1}{2}}\left\{ \sum _{x=0,1}S\left( \begin{array}{cc}
\frac{1}{2} & x\\
a & \frac{1}{2}
\end{array}\right) (\theta -i\nu _{n})S\left( \begin{array}{cc}
\frac{1}{2} & b\\
x & \frac{1}{2}
\end{array}\right) (\theta +i\nu _{n})R^{x}_{\frac{1}{2}\frac{1}{2}}(\theta )\right\} \label{01plus} \\
 &  & \mathrm{where}\quad g^{\frac{1}{2}}_{a}\equiv g^{\left| 1/2\right\rangle }_{\left| \left. a,1/2\right| n\right\rangle }\nonumber 
\end{eqnarray}
The result turns out to be\begin{equation}
\label{R_excited}
R_{ab}^{\frac{1}{2}}(\theta )=P(\theta )K(\theta +i\nu _{n})K(\theta -i\nu _{n})\frac{g_{b}^{\frac{1}{2}}}{g_{a}^{\frac{1}{2}}}\left( \delta _{ab}\cos \left( \frac{\nu _{n}}{2}\right) +\delta _{a,1-b}\sin \left( \frac{\theta }{2i}\right) \right) 
\end{equation}
Note the appearance of the \( g \) factors in the result. They are the
SUSY parts of the boundary couplings and come in two types: one
corresponds to the absorption of the particle while creating a higher
excited boundary state, the other describing the emission of the
particle and transition to some lower excited boundary state. The ones
above are of the absorption type. The residues of the full reflection
factor are described by the product of an emission and an absorption
type full boundary coupling (for a definition of boundary couplings
and their relation to the residue of the reflection factor see
\cite{GZ}). Due to the tensor product structure the full boundary
coupling is given by the bosonic part multiplied with the SUSY \( g
\)-factor, as in the case of bulk scattering \cite{hollowood}. The
product of the appropriate emission and absorption SUSY \( g
\)-factors is constrained to coincide with the value of the SUSY part
of the reflection factor at the position of the pole in the bosonic
factor. It can be seen in general that this does not give enough
constraints to determine their value unambiguously due to the
degeneracy introduced by the RSOS indices \( a,b \), and as no
physical quantity should explicitly depend on their value (see the
example of the relation \( \tilde{Q}^{2}_{+}=2\left( \tilde{H}+M\,
\tilde{Z}\right) \) discussed above) we do not present any solution
for them.

Being constructed by the bootstrap, the reflection factors (\ref{R_excited})
necessarily satisfy the constraints of boundary factorization and crossing-unitarity;
in addition, they commute with \( \tilde{Q}_{+} \) which is guaranteed by the
fact that the action of the boundary supercharge is also derived from the bootstrap
as in (\ref{excited_Q}). The full reflection factor on the \(\vert
a,1/2\vert n\rangle \) excited boundary 
can be obtained by multiplying
this result with the appropriate excited bosonic reflection factor:\begin{equation}
\label{t_excited}
R^{\frac{1}{2}}_{ab}\left( \theta \right) \times Q_{\left| n\right\rangle }\left( \eta ,\vartheta ,\theta \right) \qquad \mathrm{or}\qquad R^{\frac{1}{2}}_{ab}\left( \theta \right) \times P^{\pm }_{\left| n\right\rangle }\left( \eta ,\vartheta ,\theta \right) \; .
\end{equation}

Clearly (\ref{R_excited}) has neither pole nor zero in the physical strip.
So the poles of the reflection factors on the first excited wall (\ref{t_excited})
are exactly the same as in the non supersymmetric theory: that is they are at
\( i\nu _{k} \) or at \( iw_{m} \). 

\vspace{0.3cm}
{\par\centering \begin{figure}~~~~~~~~~~~~~~~~~~~~\subfigure[$w$ type poles]{\resizebox*{!}{5cm}{\includegraphics{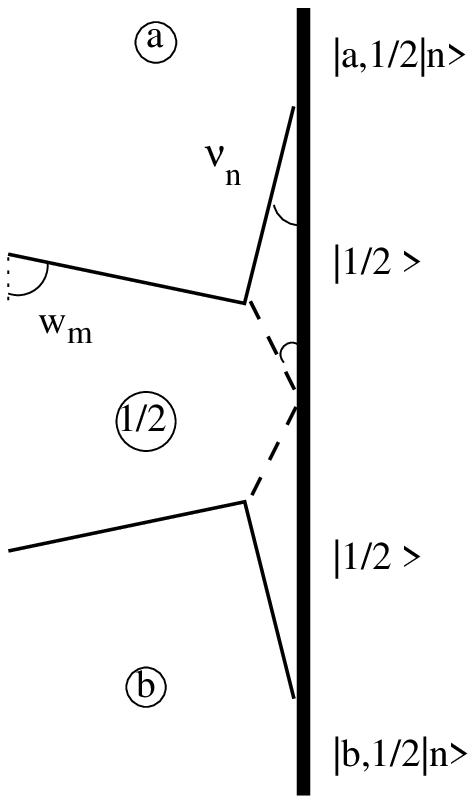}}} ~~~~~~~~~~~~~~~~~~~~~~~~\subfigure[$\nu$ type poles]{\resizebox*{!}{5cm}{\includegraphics{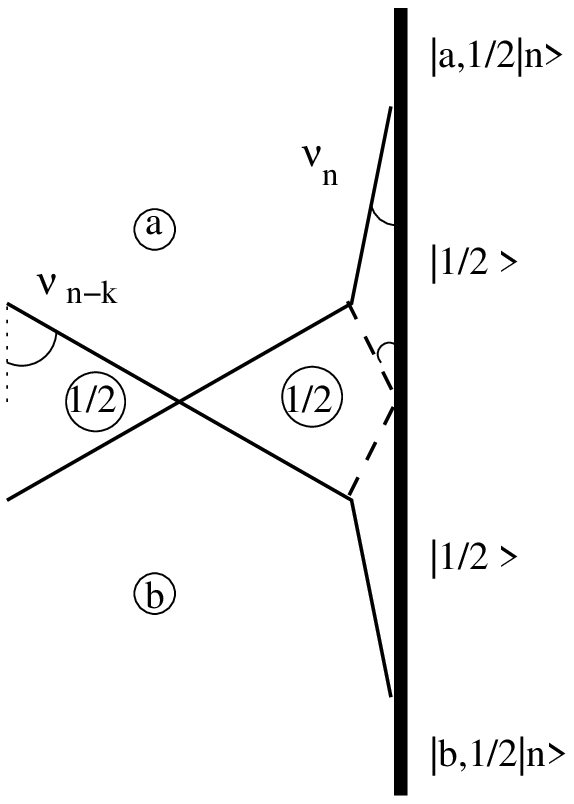}}} \end{figure}\par}
\vspace{0.3cm}

The decoration of the non supersymmetric diagrams shows, that diagram (e) explains
the \( \nu  \) type of poles, while diagram (d) explains the \( w \) type,
but only for \( w_{m}>\nu _{n} \). For \( w_{m}<\nu _{n} \) we have a boundary
bound state which we denote by \[
\left| \left. \frac{1}{2},a,\frac{1}{2}\right| m,n\right\rangle =\frac{1}{g^{\left| \left. a,1/2\right| n\right\rangle }_{\left| \left. \frac{1}{2},a,\frac{1}{2}\right| m,n\right\rangle }}K_{\frac{1}{2}a}\left( i\, w_{m}\right) \left| \left. a,1/2\right| n\right\rangle \; ,\]
 so this is also a doublet, but now it is the \( K_{a\frac{1}{2}} \) type kinks
that are able to reflect on it. It can be checked easily that the relation \( \tilde{Q}^{2}_{+}=2\left( \tilde{H}+M\, \tilde{Z}\right)  \)
holds for these states as well, consistently with the previous interpretation
of \( \gamma  \) (for these states \( \tilde{Z}=0 \)).

At this point the question emerges whether the two states \( a=0,\, 1 \) forming
the doublet \( \left| \left. \frac{1}{2},a,\frac{1}{2}\right| m,n\right\rangle  \)
are physically different or there is a possibility for some identification so
that a single state can explain the pole in the reflection matrix (\ref{t_excited}).
This can be decided by examining whether one can describe the residue of the
reflection factor with a single intermediate state, which implies a relation
between the \( R_{00}^{1/2}\, ,R_{11}^{1/2}\, ,R_{10}^{1/2} \) and \( R_{01}^{1/2} \) components of
the reflection factor at the pole. This relation is violated (for
generic values of the parameters) 
and so one must
really introduce the two states above. 

Following the same analysis we performed in \cite{BSG}, but now using a decorated
version of the Coleman-Thun diagrams it can be seen that the poles in the reflection
matrix on the above boundary excited state, which can not be explained by Coleman-Thun
diagrams are located at \( i\nu _{k} \). Since the poles appear in association
with both reflection factors \( R^{b}_{\frac{1}{2}\frac{1}{2}}(\theta ) \),
the corresponding boundary states, which are denoted by \[
\left| \left. b,\frac{1}{2},a,\frac{1}{2}\right| k,m,n\right\rangle \]
 have a fourfold degeneracy. 

It is clear that the general boundary bound state has the structure\begin{equation}
\label{rsos_excited}
\left| \left. a_{k}\dots \frac{1}{2},a_{1},\frac{1}{2}\right| n_{k}\dots ,m_{1},n_{1}\right\rangle \quad \mathrm{or}\quad \left| \left. \frac{1}{2},a_{k}\dots \frac{1}{2},a_{1},\frac{1}{2}\right| m_{k},n_{k}\dots ,m_{1},n_{1}\right\rangle 
\end{equation}
From this we see that in the supersymmetric case the boundary excited states
have a nontrivial degeneracy in contrast to the bosonic theory. The degeneracy
is labeled by RSOS sequences starting from \( 1/2 \) . In both states in (\ref{rsos_excited})
the labels \( a_{i} \) can freely take the values \( 0 \) and \( 1 \), and,
as a result, the degeneracy of the states is \( 2^{k} \). The associated reflection
factors can be computed from successive application of the bootstrap procedure,
which is illustrated on the figure

\vspace{0.3cm}
{\par\centering \resizebox*{!}{5cm}{\includegraphics{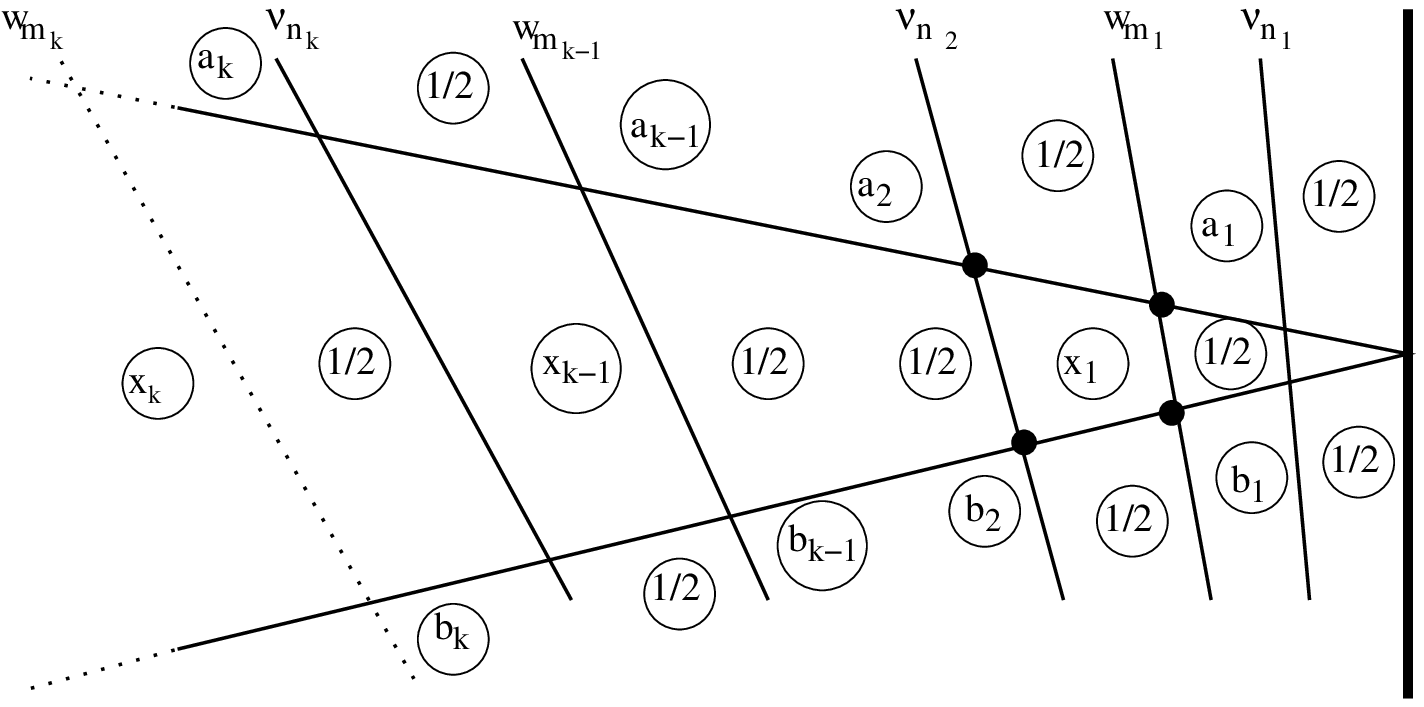}} \par}
\vspace{0.3cm}

The result depends on the \( \tilde{Z} \) charge of the scattering particles.
In the \( \tilde{Z}=0 \) case the result can be written in the following form\begin{equation}
\label{sol_excited}
R^{\frac{1}{2}}(\theta )^{\left| \left. b_{k}\dots \frac{1}{2},b_{1},\frac{1}{2}\right| n_{k}\dots ,m_{1},n_{1}\right\rangle }_{\left| \left. a_{k}\dots \frac{1}{2},a_{1},\frac{1}{2}\right| n_{k}\dots ,m_{1},n_{1}\right\rangle }=R_{a_{1}b_{1}}^{\frac{1}{2}}(\theta )\prod ^{k-1}_{i=1}f^{a_{i}a_{i+1}}_{b_{i}b_{i+1}}(w_{m_{i}},\nu _{n_{i+1}},\theta )
\end{equation}
where \( f^{a_{1}a_{2}}_{b_{1}b_{2}}(w_{m_{1}},\nu _{n_{2}},\theta ) \) is
the contribution of the dotted square summing over \( x_{1}=0,1 \) that is\begin{eqnarray*}
f^{a_{1}a_{2}}_{b_{1}b_{2}}(w_{m_{1}},\nu _{n_{2}},\theta ) & = & \sum _{x_{1}=0,1}S\left( \begin{array}{cc}
x_{1} & \frac{1}{2}\\
\frac{1}{2} & a_{1}
\end{array}\right) (\theta -iw_{m_{1}})S\left( \begin{array}{cc}
x_{1} & \frac{1}{2}\\
\frac{1}{2} & b_{1}
\end{array}\right) (\theta +iw_{m_{1}})\\
 &  & \\
 & \times  & S\left( \begin{array}{cc}
\frac{1}{2} & x_{1}\\
a_{2} & \frac{1}{2}
\end{array}\right) (\theta -i\nu _{n_{2}})S\left( \begin{array}{cc}
\frac{1}{2} & b_{2}\\
x_{1} & \frac{1}{2}
\end{array}\right) (\theta +i\nu _{n_{2}})
\end{eqnarray*}
Collecting the common factors we have\begin{eqnarray*}
f^{a_{1}a_{2}}_{b_{1}b_{2}}(w_{m_{1}},\nu _{n_{2}},\theta )=K(\theta -iw_{m_{1}})K(\theta +iw_{m_{1}})K(\theta -i\nu _{n_{2}})K(\theta +i\nu _{n_{2}}) &  & \\
\times\frac{g^{|a_{1},\frac{1}{2}|n_{1}\rangle }_{|\frac{1}{2}a_{1},\frac{1}{2}|m_{1},n_{1}\rangle }g_{|a_{2},\frac{1}{2}a_{1},\frac{1}{2}|n_{2},m_{1},n_{1}\rangle }^{|\frac{1}{2}a_{1},\frac{1}{2}|m_{1},n_{1}\rangle }}{\left\{ a\leftrightarrow b\right\} }h^{a_{1}a_{2}}_{b_{1}b_{2}}(w_{m_{1}},\nu _{n_{2}},\theta ) &  & 
\end{eqnarray*}
 where\begin{eqnarray*}
h^{a_{1}a_{2}}_{b_{1}b_{2}}(w_{m_{1}},\nu _{n_{2}},\theta )=2\sum _{x=0,1}\cos \left( \frac{\theta -iw_{m_{1}}}{4i}+\frac{\pi }{4}(1-2\delta _{x,a_{1}})\right) \cos \left( \frac{\theta +iw_{m_{1}}}{4i}+\frac{\pi }{4}(1-2\delta _{x,b_{1}})\right)  &  & \\
\times\cos \left( \frac{\theta -i\nu _{n_{2}}}{4i}-\frac{\pi }{2}\delta _{x,a_{2}})\right) \cos \left( \frac{\theta +i\nu _{n_{2}}}{4i}-\frac{\pi }{2}(1-\delta _{x,b_{2}})\right)  &  & 
\end{eqnarray*}
 In the \( \tilde{Z}=1 \) case, which is indicated with dotted lines on the
diagram, the result contains an extra factor\begin{equation}
\label{sol_excited1}
R^{x_{k}}(\theta )^{\left| \left. \frac{1}{2},b_{k}\dots \frac{1}{2},b_{1},\frac{1}{2}\right| m_{k}\dots ,m_{1},n_{1}\right\rangle }_{\left| \left. \frac{1}{2},a_{k}\dots \frac{1}{2},a_{1},\frac{1}{2}\right| m_{k}\dots ,m_{1},n_{1}\right\rangle }=h^{x_{k}}_{a_{k},b_{k}}R^{\frac{1}{2}}(\theta )^{\left| \left. b_{k}\dots \frac{1}{2},b_{1},\frac{1}{2}\right| n_{k}\dots ,m_{1},n_{1}\right\rangle }_{\left| \left. a_{k}\dots \frac{1}{2},a_{1},\frac{1}{2}\right| n_{k}\dots ,m_{1},n_{1}\right\rangle }
\end{equation}
 where 
\[
h^{x_{k}}_{a_{k},b_{k}}=2^{-\frac{\theta }{i\pi }} 
\frac
{g^{\left|a_k\frac{1}{2}\dots\right\rangle}_{\left|\frac{1}{2}a_{k-1}\dots\right\rangle}}
{g^{\left|b_k\frac{1}{2}\dots\right\rangle}_{\left|\frac{1}{2}b_{k-1}\dots\right\rangle}}
\left( -\sin \left( \frac{\theta }{2i}-\frac{\pi }{2}(\delta _{x_{k},a_{k}}+\delta _{x_{k},b_{k}})\right) +\cos \left( \frac{w_{m_{k}}}{2}+\frac{\pi }{2}(\delta _{x_{k},a_{k}}-\delta _{x_{k},b_{k}})\right) \right) \]

\subsubsection{The \protect\( \Gamma \protect \) non symmetric (\protect\( BSSG^{-}\protect \))
case}

The discussion of the \( \Gamma  \) non symmetric (\( BSSG^{-} \)) solution
runs entirely parallel to the previous \( \Gamma  \) symmetric case, the only
difference being that in the input of the bootstrap procedure, i.e. in the ground
state reflection amplitude \[
R^{a}_{\frac{1}{2}\frac{1}{2}}(\theta )\times R_{SG}(\theta )\]
 the supersymmetry factors, \( R^{a}_{\frac{1}{2}\frac{1}{2}}(\theta ) \),
\( a=0,1 \), are taken now from (\ref{Rminus}). These factors depend explicitly
on \( \gamma  \), and this dependence pertains in the (SUSY) reflection amplitudes
on exited boundaries. Nevertheless, since none of these amplitudes has a pole
in the physical strip, following the steps of the previous considerations leads
to the same conclusion regarding the indexing and degeneracies of the boundary
states. Therefore we concentrate here mainly on the differences between the
two solutions.

At the position of the \( \nu _{n} \) poles in the ground state reflection
amplitude we again associate boundary bound states \( |a,1/2|n\rangle  \) to
\( R^{a}_{\frac{1}{2}\frac{1}{2}} \) \( a=0,1 \) as in (\ref{first_fusion})
(though of course the present values of boundary couplings may differ from the
previous ones). The action of the present boundary supercharge, \( \tilde{Q}_{-} \),
on these states is \[
\tilde{Q}_{-}|0,1/2|n\rangle =r\left( \gamma +2\sqrt{M}\sin \frac{\nu _{n}}{2}\right) |1,1/2|n\rangle ,\]
 \[
\tilde{Q}_{-}|1,1/2|n\rangle =r^{-1}\left( \gamma -2\sqrt{M}\sin \frac{\nu _{n}}{2}\right) |0,1/2|n\rangle .\]
 The action of \( \tilde{Q}_{-}^{2} \) on these states is compatible with the
relation \( \tilde{Q}_{-}^{2}=2(\tilde{H}-M\tilde{Z}) \), provided we keep
the interpretation of \( \gamma ^{2}/2 \) as the ground state energy.

Using the SUSY factors eqn. (\ref{Rminus}) in the bootstrap equation (\ref{01plus})
for the reflections of the \( K_{\frac{1}{2}a} \) kinks on these boundary states
gives \begin{eqnarray*}
R^{\frac{1}{2}}_{ab}(\theta ) & = & Z_{-}(\theta )\frac{g_{b}^{\frac{1}{2}}}{g_{a}^{\frac{1}{2}}}\Bigg [\delta _{ab}\left( \cos \frac{\xi }{2}\cos \frac{\nu _{n}}{2}+(-1)^{a}i\sinh \frac{\theta }{2}\cosh \frac{\theta }{2}\right) \\
 & - & i\delta _{a,1-b}\sinh \frac{\theta }{2}\left( \cos \frac{\xi }{2}+(-1)^{b}\sin \frac{\nu _{n}}{2}\right) \Bigg ]\; ,
\end{eqnarray*}
 where \( \xi  \) is expressed in terms of \( \gamma  \) in eqn. (\ref{gamma_xi}),
and \[
Z_{-}(\theta )=P(\theta )K(\theta +i\xi )K(\theta -i\xi )K(\theta +i\nu _{n})K(\theta -i\nu _{n}).\]
Although this reflection amplitude has a slightly more complicated form than
the one in eqn. (\ref{R_excited}), it also solves the boundary Yang-Baxter
equation and the other constraints by construction. The difference between the
two comes from the fact that (\ref{R_excited}) commutes with \( \tilde{Q}_{+} \),
while the present reflection factor is invariant under \( \tilde{Q}_{-} \).

Finally we point out that the expressions (\ref{sol_excited},\ref{sol_excited1})
for the kink reflection amplitudes on the general higher excited boundary states
remain valid in the case \( BSSG^{-} \) as well, if the ground state reflection
factor (\ref{Rplus}) is replaced by (\ref{Rminus}).

\section{Breather reflection factors}

The breathers have vertex type scattering matrices in contrast to the RSOS type
ones of the kinks. These scattering matrices enter into the equations determining
the reflection factors of the breathers, nevertheless there is no need for their
explicit form as the breather reflection factors on the various boundaries can
be obtained from that of the soliton kinks by using the (bulk) fusion and the
bootstrap \cite{bulk_bootstrap}; the procedure is summarized schematically
on diagram (f). 

\vspace{0.3cm}
{\par\centering \begin{figure}~~~~~~~~~~\subfigure[The bootstrap procedure for the breather reflection factors on the boundary ground state $\left|\frac{1}{2}\right\rangle$]{\includegraphics{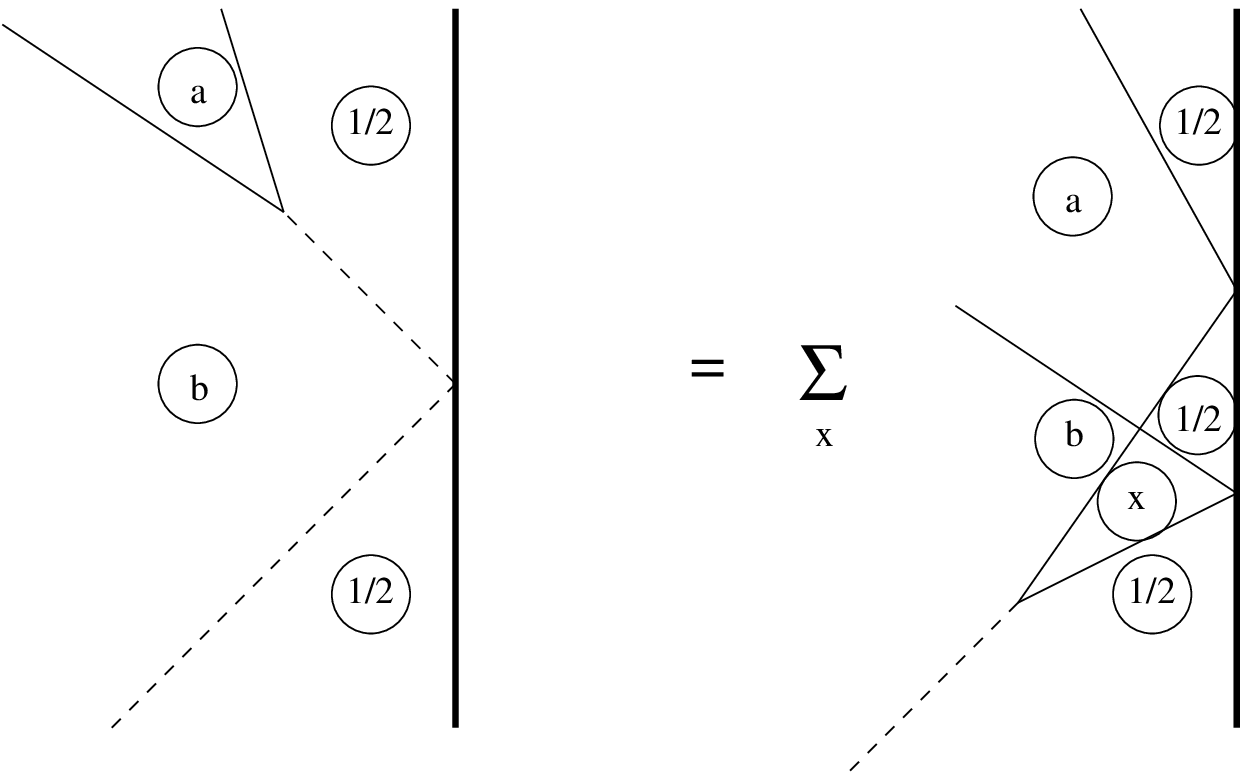}} \end{figure}\par}
\vspace{0.3cm}

If two bulk kinks form a bound state at a rapidity difference \( i\rho  \)
(\( 0<\rho <\pi  \)) the bound state is identified with a supermultiplet \( (\phi ,\, \, \psi ) \)
of mass \( 2M\cos (\rho /2) \). (In case of the \( k \)th breather \( \rho =\rho _{k}=\pi -\frac{k\pi }{\lambda } \)).
The fusing coefficients of these processes are defined via \cite{hollowood}:
\[
|K_{ab}(\theta +i\rho /2)K_{bc}(\theta -i\rho /2)\rangle ={f^{\phi }}_{abc}|\phi (\theta )\rangle +{f^{\psi }}_{abc}|\psi (\theta )\rangle \]
 with the non vanishing coefficients being: \[
{f^{\phi }}_{0\frac{1}{2}0}={f^{\phi }}_{1\frac{1}{2}1}=2^{(\pi -2\rho )/4\pi }{f^{\phi }}_{\frac{1}{2}0\frac{1}{2}}=2^{(\pi -2\rho )/4\pi }{f^{\phi }}_{\frac{1}{2}1\frac{1}{2}}=\sqrt{K(i\rho )2^{(\pi -\rho )/2\pi }\cos \left( \frac{\rho -\pi }{4}\right) }\]
 and \[
{f^{\psi }}_{1\frac{1}{2}0}=-{f^{\psi }}_{0\frac{1}{2}1}=2^{(\pi -2\rho )/4\pi }i{f^{\psi }}_{\frac{1}{2}0\frac{1}{2}}=-2^{(\pi -2\rho )/4\pi }i{f^{\psi }}_{\frac{1}{2}1\frac{1}{2}}=\sqrt{K(i\rho )2^{(\pi -\rho )/2\pi }\cos \left( \frac{\rho +\pi }{4}\right) }\]
To describe the ground state reflection amplitudes of the bosonic (\( \phi  \))
and fermionic (\( \psi  \)) components we represent them as \[
\phi (\theta )|\frac{1}{2}\rangle =\frac{1}{2{f^{\phi }}_{\frac{1}{2}0\frac{1}{2}}}\left( K_{\frac{1}{2}0}(\theta +i\rho /2)K_{0\frac{1}{2}}(\theta -i\rho /2)+K_{\frac{1}{2}1}(\theta +i\rho /2)K_{1\frac{1}{2}}(\theta -i\rho /2)\right) |\frac{1}{2}\rangle ,\]
 \[
\psi (\theta )|\frac{1}{2}\rangle =\frac{1}{2{f^{\psi }}_{\frac{1}{2}0\frac{1}{2}}}\left( K_{\frac{1}{2}0}(\theta +i\rho /2)K_{0\frac{1}{2}}(\theta -i\rho /2)-K_{\frac{1}{2}1}(\theta +i\rho /2)K_{1\frac{1}{2}}(\theta -i\rho /2)\right) |\frac{1}{2}\rangle .\]
 These expressions show that they also provide an ordinary doublet representation
of the boundary supercharge \( \tilde{Q}_{\pm } \) and that the fermionic parity
\( \Gamma  \) act on them in the standard way. The actual reflection factors
are obtained from the bootstrap equation on diagram (f), where the dashed lines
represent either \( \phi  \) or \( \psi  \). The bosonic and fermionic reflection
factors are qualitatively different in the \( \Gamma  \) symmetric and \( \Gamma  \)
non symmetric cases, since the bootstrap equations contain both the \( R^{0}_{\frac{1}{2}\frac{1}{2}} \)
and the \( R^{1}_{\frac{1}{2}\frac{1}{2}} \) ground state kink reflection amplitudes,
and these are significantly different in the two cases. Writing the breather
reflection factors on the ground state boundary as \[
\left( \begin{array}{c}
\phi \\
\psi 
\end{array}\right) (\theta )|\frac{1}{2}\rangle =\left( \begin{array}{ll}
\mathcal{A}_{+} & \mathcal{B}\\
\tilde{\mathcal{B}} & \mathcal{A}_{-}
\end{array}\right) \left( \begin{array}{c}
\phi \\
\psi 
\end{array}\right) (-\theta )|\frac{1}{2}\rangle \]
 in the \( \Gamma  \) symmetric (\( BSSG^{+} \)) case one obtains \begin{eqnarray*}
{\mathcal{B}}={\tilde{\mathcal{B}}}=0,\qquad {\mathcal{A}}_{+}=Z(\theta )\cos \left( \frac{\theta }{2i}-\frac{\pi }{4}\right) ,\qquad {\mathcal{A}}_{-}=Z(\theta )\cos \left( \frac{\theta }{2i}+\frac{\pi }{4}\right) \; , &  & \\
Z(\theta )=P(\theta +i\rho /2)P(\theta -i\rho /2)\sqrt{2}K(2\theta )2^{-\theta /i\pi }\; , &  & 
\end{eqnarray*}
 while in the \( \Gamma  \) non symmetric (\( BSSG^{-} \)) case we get \[
{\mathcal{B}}=-{\tilde{\mathcal{B}}}=-\tilde{Z}(\theta )\frac{\gamma }{2\sqrt{M}}\sqrt{\cos (\rho /2)}\sinh (\theta ),\]
 \begin{eqnarray*}
{\mathcal{A}}_{\pm }=\tilde{Z}(\theta )\Bigg (\cosh \left( \frac{\theta }{2}\right) \left( \frac{\gamma ^{2}}{4M}-\left[ \sin ^{2}\left( \frac{\rho }{4}\right) +\sinh ^{2}\left( \frac{\theta }{2}\right) \right] \right)  &  & \\
\mp i\sinh \left( \frac{\theta }{2}\right) \left( \frac{\gamma ^{2}}{4M}+\left[ \sin ^{2}\left( \frac{\rho }{4}\right) +\sinh ^{2}\left( \frac{\theta }{2}\right) \right] \right) \Bigg )\, , &  & 
\end{eqnarray*}
 with \[
\tilde{Z}(\theta )=K(2\theta )2^{-\theta /i\pi }F(\theta -i\rho /2)F(\theta +i\rho /2),\quad F(\theta )=P(\theta )K(\theta +i\xi )K(\theta -i\xi )\: .\]
The two cases are indeed qualitatively different: in the \( BSSG^{+}
\) solution the conservation of fermionic parity forbids the \( \phi
\rightarrow \psi \) reflection on the ground state boundary, while in
the \( BSSG^{-} \) solution this reflection is possible. The form of
the \( BSSG^{+} \) solution is the same as the one obtained in
\cite{moriconi_schoutens} by imposing fermion number conservation. The
structural form of the reflection factors in the case \( BSSG^{-} \)
are identical to the ones obtained in \cite{ahn_koo2} from the eight
vertex free fermion model with boundary. In \cite{nepomechie} it was
proposed that $\gamma$, which appears explicitly in the reflection
matrix, could be fixed in terms of $\eta$ and $\vartheta$ by
the boundary bootstrap. Here we see that this is not the case, as the
bootstrap gives no constraint for $\gamma$, due to degeneracies appearing in the
boundary excited states. We recall however that $\gamma$ is not a free
parameter, as it is determined by the ground state boundary energy (or
equivalently, by the boundary supercharge), thus it can be expressed
in terms of the Lagrangian parameters in principle. 

A more explicit description of the reflection factor of the first
breather on the ground state boundary in the $BSSG^-$ case was given
in boundary sinh-Gordon model studied in \cite{ahn_nepomechie}, but
the precise connection between the parameters used in that paper and
the present one is yet to be determined.

In the bosonic theory the ground state reflection amplitude of the \( k \)th
breather has poles at \[
-i\theta =\frac{\eta }{\lambda }-\frac{\pi }{2}+(k-2l-1)\frac{\pi }{2\lambda }\quad ,\quad l=0,\dots ,\, \left[ \frac{k-1}{2}\right] \; .\]
In the supersymmetric theory, these poles signal the presence of the excited
boundary states\[
\left| \left. \frac{1}{2},a,\frac{1}{2}\right| l,k-l\right\rangle \]
as intermediate states in the breather reflection process. Since there are two
intermediate states (\( a=0,1 \)), the determinant of the \( 2\times 2 \)
reflection matrix should not vanish at the position of these poles (as the residue
of the reflection matrix at the pole is proportional to the projector on the
subspace of on-shell intermediate states). It is straightforward to verify that
this is indeed the case for both the \( BSSG^{+} \) and the \( BSSG^{-} \)
solutions.

Using the bootstrap procedure it is also possible to obtain the breather reflection
factors on excited boundaries. If the RSOS sequence characterizing the boundary
state ends with a label \( a \) (\( a=0 \) or \( a=1 \)), then e.g. the bosonic
breather can be represented by 
\[
\left|\left.\phi (\theta )a,\frac{1}{2},\dots \right|n_{1},\dots \right\rangle =\frac{1}{{f^{\phi }}_{a\frac{1}{2}a}}K_{a\frac{1}{2}}(\theta +i\rho /2)K_{\frac{1}{2}a}(\theta -i\rho /2)\left|\left.a,\frac{1}{2},\dots \right|n_{1},\dots \right\rangle \]
 in the bootstrap procedure. To emphasize that even in the \( BSSG^{+} \) case
there are \( \phi \, \rightarrow \, \psi  \) type reflections on excited boundaries
we give here the reflection matrix of breathers on the \( |a1/2|n\rangle  \)
states. In the basis of \( |\phi (\theta )0, 1/2|n\rangle  \), \( |\phi (\theta )1, 1/2|n\rangle  \),
\( |\psi (\theta )0, 1/2|n\rangle  \) \( |\psi (\theta )1, 1/2|n\rangle  \)
it can be written as\begin{equation}
\label{bref_matrix}
\hat{Z}(\theta )\left( \begin{array}{cccc}
\mathcal{C}_{+} & 0 & 0 & -\mathcal{D}/r\\
0 & \mathcal{C}_{+} & r\mathcal{D} & 0\\
0 & \mathcal{D}/r & \mathcal{C}_{-} & 0\\
-r\mathcal{D} & 0 & 0 & \mathcal{C}_{-}
\end{array}\right) 
\end{equation}
 where \( {\mathcal{D}}=\frac{1}{\sqrt{2}}\sqrt{\cos \frac{\rho }{2}}\cos \frac{\nu _{n}}{2}\sin \frac{\theta }{i} \)
and \[
{\mathcal{C}}_{\pm }=\cos ^{2}\frac{\nu _{n}}{2}\cos \left( \frac{\theta }{2i}\mp \frac{\pi }{4}\right) +\frac{1}{2}\left( \cos \frac{\rho }{2}-\cos \frac{\theta }{i}\right) \cos \left( \frac{\theta }{2i}\pm \frac{\pi }{4}\right) \, .\]
 It is easy to show that in spite of the non trivial boson fermion reflection
the operator \( \Gamma  \) commutes with this reflection matrix. 

A nontrivial check on the consistency of the bootstrap solution can be obtained
by considering the pole structure of the full reflection amplitude
containing the SUSY factor (\ref{bref_matrix}). The bosonic reflection
factor of \( \mathbf{B}^{k} \) on the bosonic boundary excited state \( \left| n\right\rangle  \)
has a pole at \( -i\theta =\frac{\pi }{2}-\frac{\eta }{\lambda }+\frac{\pi }{2\lambda }\left( k+2n+1\right)  \)
\cite{BSG}. In the supersymmetric case, this means that a boundary excited
state of the form\begin{equation}
\label{fusion_process}
\left| \left. a,\frac{1}{2}\right| n+k\right\rangle 
\end{equation}
enters as an on-shell intermediate state in the scattering of \( \mathbf{B}^{k} \)
on \( |a,1/2|n\rangle  \). However, due to the doublet (boson/fermion) structure
of the breather naively one would expect \( 4 \) states to explain the residue
of the \( 4\times 4 \) reflection factor. In the conjectured spectrum, on the
other hand, the only possible process goes via (\ref{fusion_process}) and it
allows for only two intermediate states (\( a=0,\, 1 \)). Therefore one expects
that the determinant of the matrix (\ref{bref_matrix}) should have a double
zero there. It can be verified by direct calculation that this double
zero is indeed there without imposing any restriction on the parameters.

In the reflection of the \( k \)th breather on the \( n \)th excited boundary,
there is another family of poles at \( -i\theta =\frac{\eta }{\lambda }-\frac{\pi }{2\lambda }\left( k-2l+1\right) \; ,\; l=0,\dots ,n-1 \)
\cite{BSG} that in the supersymmetric case should correspond to intermediate
states of the form\[
\left| \left. b,\frac{1}{2},a,\frac{1}{2}\right| l,k-l,n\right\rangle \: .\]
At these poles, the number of intermediate states is \( 4 \) (\( a,b=0,1 \))
and so we expect that the determinant of the SUSY factor does not vanish, which
indeed turns out to be the case.

\section{Conclusions}

To start we summarize the results of this paper. We considered the boundary
scattering amplitudes in boundary supersymmetric sine-Gordon theory (\( BSSG \)).
Imposing the constraint of supersymmetry on solutions of the boundary Yang-Baxter
equation, we found two consistent sets of amplitudes that describe the reflection
of solitons off the boundary in its ground state. Then we considered the two
bootstrap systems built from these fundamental amplitudes and conjectured the
closure of this bootstrap , i.e. the set of boundary states and the reflection
factors on them. We also derived a relation between the boundary supercharge
and the Hamiltonian and checked that this relation holds for the bootstrap solutions. 

Although the reflection amplitudes are different, the spectrum of states is
the same in the two bootstrap solutions. This common spectrum is characterized
partly by a sequence of integers, just like in the case of the ordinary sine-Gordon
model \cite{BSG}, but also by an RSOS sequence of length \( k+1 \) (if the
length of the integer sequence is \( k \)) starting from \( 1/2 \). The energy
of the state depends only on the integer labels, the different RSOS sequences
correspond to degenerate states. It is interesting to note that the non supersymmetric
boundary spectrum allows for a tensor product type supersymmetrization, and
no further constraints are obtained in accord with the bulk case \cite{hollowood}. 

In the case of the \( BSSG^{+} \) theory, the reflection amplitudes depend
on two parameters \( \eta  \) and \( \vartheta  \), that are inherited from
the bosonic reflection factors and were originally introduced in \cite{GZ}.
In the bosonic case it is known how these parameters are related to the parameters
of the boundary Lagrangian in the perturbed CFT formalism \cite{uv_ir}. Besides
that, the SUSY algebra introduces a further parameter \( \gamma  \), which
is related to the energy of the boundary ground state and so must be a function
of the parameters of the \( BSSG \) Lagrangian. In the \( BSSG^{-} \) theory
the difference is that \( \gamma  \) appears also in the expression for the
reflection factors themselves. In the bosonic case the expression for the boundary
energy in terms of Lagrangian parameters is also known \cite{uv_ir}. The existence
of two different families of solutions and the number of parameters are in accordance
with the expectations that they describe the scattering in the Lagrangian theories
corresponding to the boundary interaction (\ref{boundary_interaction})
\cite{nepomechie}.

It is a very interesting and important issue to connect the bootstrap parameters
\( \eta ,\, \theta  \) and the vacuum energy parameter \( \gamma  \) to the
parameters of the Lagrangian description for the supersymmetric case
as well.  In the case of the non supersymmetric boundary sine-Gordon
theory this was achieved by considering it as a combined bulk and
boundary perturbation of a $c=1$ free massless boson with Neumann
boundary condition. However, even the interpretation of the bulk \(
SSG \) theory as a perturbed CFT is nontrivial, and we are
investigating this problem. We are also working on getting more
evidence to link the bulk
\( S \) matrix and the reflection factors to the Lagrangian theory. Work is
in progress in these directions and we hope to report on the results in the very
near future.

\subsubsection*{Acknowledgments}

The authors would like to thank G.M.T. Watts for very useful
discussions and R.I. Nepomechie for comments on the manuscript. G.T.
thanks the Hungarian Ministry of Education for a Magyary Postdoctoral
Fellowship, while B.Z. acknowledges partial support from a Bolyai
János Research Fellowship.
This research was supported in part by the Hungarian Ministry of Education under
FKFP 0043/2001 and the Hungarian National Science Fund (OTKA) grants T037674/02
and T34299/01.

\end{document}